\documentclass[prd,aps]{revtex4}

\usepackage{epsfig}

\newcommand{\beq}{\begin{equation}}
\newcommand{\eeq}{\end{equation}}
\newcommand{\be}{\begin{eqnarray}}
\newcommand{\ee}{\end{eqnarray}}

\newcommand{\bs}{B_s-\bar{B}_s}

\begin{document}

\preprint{
\begin{tabular}{l}
\hbox to\hsize{June, 2008\hfill KIAS-P08040}\\[-2mm]
\end{tabular}
}

\title{
Implications of the anomalous top quark couplings 
\\
in $B_s-\bar{B}_s$ mixing, $B \to X_s \gamma$ and top quark decays
}

\author{ Jong Phil Lee }

\affiliation{
School of Physics, Korea Institute for Advanced Study, Seoul 130-722, Korea
}

\author{ Kang Young Lee }
\email{kylee@muon.kaist.ac.kr}

\affiliation{
Department of Physics, Korea University, Seoul 136-713, Korea 
}

\date{\today}

\begin{abstract}

Combined analysis of recent measured $\bs$ mixing 
and $B \to X_s \gamma$ decays
provides constraints on the anomalous $\bar{t}sW$ couplings.
We discuss the perspectives to examine
the anomalous $\bar{t}sW$ couplings
through CKM-suppressed $ t \to s W$ decays at the LHC.

\end{abstract}

\maketitle



\section{Introduction}

The standard model (SM) has been demonstrated to be remarkably
successful in describing present data.
Most parameters of the SM has been directly measured
with high accuracy at various experiments.
The only unobserved ingredient of the SM is the Higgs boson
responsible for the electroweak symmetry breaking
and a few top quark couplings are not measured directly.
However, it is not believed that the SM is the final theory 
of our universe since there are still many theoretical
and experimental problems which could not be explained
in the SM framwork.
It is natural to expect that the hint of the new physics
beyond the SM would be found at the unexamined part of the SM.

The top quark has been discovered at the Tevatron 
and its mass and production cross section are measured \cite{top}.
We will be able to study the top quark couplings
with more than $10^8$ top quark pairs per year 
produced at the CERN Large Hadron Collider (LHC)
\cite{lhc,toplhc}.
The dominant channel of the top quark decay is 
the $t \to b W$ channel in the SM and
the $\bar{t}bW$ coupling will be measured at LHC
with high precision to be directly tested.
Other channels are highly suppressed by small mixing angles.
The subdominant channel in the SM is
the Cabibbo-Kobayashi-Maskawa (CKM) nondiagonal $t \to s W$ decay
of which branching ratio is estimated as
\be
Br(t \to s W) \sim 1.6 \times 10^{-3},
\ee
when $|V_{ts}| = 0.04$ is assumed in the SM.
Although the branching ratio of this channel is rather small,
the $t \to s W$ process may be detectable at the LHC
due to the large number of top quark production
and the $\bar{t}sW$ coupling be measured to provide 
a clue to new physics beyond the SM.
Therefore the anomalous $\bar{t}sW$ coupling is worth examining at present.
We do not specify the underlying model here 
but present an effective lagrangian to describe 
the new effects on the top quark couplings
by introducing two parameters for each flavour.
The relevant couplings are parametrized by
the effective lagrangian as
\be
{\cal L} = -\frac{g}{\sqrt{2}} 
           \sum_{q=d,s,b}
          V_{tq}^{\rm eff}~ \bar{t} \gamma^\mu
                                 (P_L + \xi_q P_R) q W^+_\mu
  + H.c.,
\ee
where $\xi_q$ are complex parameters measuring effects 
of the anomalous right-handed couplings
while $V_{tq}^{\rm eff}$ measures the SM-like left-handed couplings.
Effects of the anomalous top quark couplings 
have been studied in direct and indirect ways
\cite{larios,leesong,lee2,lee,jplee,boos,rindani,elhady,yue,rizzo}.

Particularly interesting is $b \to s$ transition in search of
the anomalous top quark couplings.
The radiative decay $B \to X_s \gamma$ is the first observation
of $b \to s $ transition and provide strict constraints
on the anomalous top quark couplings
\cite{larios,leesong}
Since no CP phase is involved in $V_{ts}$ and $V_{tb}$ in the SM,
a large direct CP violation in $b \to s$
is an evidence of the new physics beyond the SM
\cite{lee2,elhady}.
Recently the first observation of the $\bs$ mixing
have been reported by the CDF \cite{cdf} and D0 \cite{d0} collaborations 
with the results
\be
&\Delta M_s 
= ( 17.77 \pm 0.10 \pm 0.07 )~ {\rm ps}^{-1}& 
~~~~~~~~~~~~~~~~~~~({\rm CDF}),
\nonumber \\
&17~ {\rm ps}^{-1} \le \Delta M_s \le 21~ {\rm ps}^{-1}& 
{\rm at}~~ 90 \%~ {\rm C.L.}~~~~
({\rm D0}),
\ee
where the first error is statistical and the second is systematic. 
The $\bs$ mixing arises through the box diagram
with internal lines of $W$ boson and $u$-type quarks in the SM.
Since the top quark loop dominates 
the $\bs$ mixing might be also a testing laboratory
for the study of the $\bar{t} s W$ and $\bar{t} b W$ couplings.

In this work, we concentrate on $\bar{t} s W$ coupling
and perform the combined analysis of
$\bs$ mixing and $B \to X_s \gamma$ to constrain
the $V_{ts}^{\rm eff}$ and $\xi_s$.
$\bs$ mixing depends upon $V_{ts}^{\rm eff}$ and is insensitive
to the right-handed couplings 
while $B \to X_s \gamma$ decay depende upon both of
$V_{ts}^{\rm eff}$ and $\xi_s$.
If we measure the subdominant decay $t \to s W$ at the LHC 
or other future colliders, it will be the direct test
of the CKM matrix element $V_{ts}^{\rm eff}$
and we can determine the $\bar{t} s W$ couplings.
This paper is organized as follows:
In section II, 
the effective $\Delta B =1$ Hamiltonian formalism 
with anomalous $\bar{t} s W$ couplings is given
and the radiative $B \to X_s \gamma$ decays are studied.
In section III, 
the analysis on the $\bs$ mixing 
with anomalous $\bar{t} s W$ couplings is presented 
We disduss the top quark decays in section IV.
Finally we conclude in section IV.

\section{$B \to X_s \gamma$} 

The $\Delta B =1$ effective Hamiltonian for
$b \to s \gamma$ process is given by 
\be
{\cal H}_{eff}^{\Delta B=1} = -\frac{4 G_F}{\sqrt{2}} V_{ts}^* V_{tb} 
           \sum_{i=1}^{8}
             \left( C_i(\mu) O_i(\mu) + C'_i(\mu) O'_i(\mu) \right),
\ee
where the dimension 6 operators $O_i$ constructed in the SM
are given in the Ref. \cite{buras},
and $O'_i$ are their chiral conjugate operators.
Matching the effective theory (5) and the lagrangian (4)
at $\mu = m_W$ scale,
we have the Wilson coefficients $C_i(\mu=m_W)$ and $C'_i(\mu=m_W)$.
Although we will consider the anomalous $\bar{t} s W$ couplings only,
we present the full formalism including
$\bar{t} s W$ and $\bar{t} b W$ couplings.
In the SM, we have the Wilson coefficients 
\be
C_2(m_W) &=& -1,~~~
C_7(m_W) = F(x_t),~~~
C_8(m_W) = G(x_t),
\nonumber \\
C_i(m_W) &=& C'_i(m_W) = 0, ~~~~~~{\rm otherwise,}
\ee
where $F(x)$ and $G(x)$ are the well-known
Inami-Lim loop functions \cite{buras,inami}.
Let us switch on the right-handed $\bar{t} b W$
and $\bar{t} s W$ couplings.
Keeping the effects of $\xi_q$ in linear order,
we obtain the modified Wilson coefficients
\be
C_7 &\to& C_7^{{\rm SM}} + \xi_b \frac{m_t}{m_b} F_R(x_t),
\nonumber \\
C_8 &\to& C_8^{{\rm SM}} + \xi_b \frac{m_t}{m_b} G_R(x_t),
\ee
and the new Wilson coefficients
\be
C'_7 &=& \xi_s \frac{m_t}{m_b} F_R(x_t),
\nonumber \\
C'_8 &=& \xi_s \frac{m_t}{m_b} G_R(x_t),
\ee
where the new loop functions 
\be
F_R(x) &=& \frac{-20+31x-5x^2}{12(x-1)^2}
                 + \frac{x (2-3x)}{2(x-1)^3} \ln x,
\nonumber \\
G_R(x) &=& -\frac{4+x+x^2}{4(x-1)^2}
                 + \frac{3x}{2(x-1)^3} \ln x,
\ee
agree with those in Ref. \cite{cho}.

The branching ratio of $B \to X_s \gamma$ process
with the right-handed interactions at next-leading-order (NLO) 
is given by
\be
{\rm Br}(B \to X_s \gamma) &=& \frac{{\rm Br}(B \to X_c e \bar{\nu})}
                                    {10.5 \%}
\left[ B_{22}(\delta) + B_{77}(\delta) (|r_7|^2 + |r'_7|^2 )
+ B_{88}(\delta) (|r_8|^2 + |r'_8|^2 )
\right.
\nonumber \\
&&~~~~~~ \left.+ B_{27}(\delta) Re (r_7)
+ B_{28}(\delta) Re (r_8)
+ B_{78}(\delta) ( Re (r_7 r^{\star}_8) + Re (r'_7 r'^{\star}_8) )
\right],
\ee
where the ratios $r_i$ and $r'_i$ are defined by
\be
r_i = \frac{C_i(m_W)}{C_i^{SM}(m_W)}
    = 1 + \xi_b \frac{m_t}{m_b} \frac{F_R(x_t)}{F(x_t)},
~~~~~~~~
r'_i =  \xi_s \frac{m_t}{m_b} \frac{F_R(x_t)}{F(x_t)},
\ee
The components $B_{ij}(\delta)$ depends on the kinematic cut $\delta$,
of which numerical values are given in the Ref. \cite{kagan}.
We obtain the branching ratio in terms of $\xi_s$ and $\xi_b$ as
\be
{\rm Br}(B \to X_s \gamma) &=& {\rm Br}^{\rm SM}(B \to X_s \gamma)
     \left( \frac{|{V_{ts}^{\rm eff}}^* V_{tb}^{\rm eff}|}
                 {0.0404} \right)^2
     \left[ 1 + Re (\xi_b) \frac{m_t}{m_b} \left(
       0.68 \frac{F_R(x_t)}{F(x_t)} + 0.07 \frac{G_R(x_t)}{G(x_t)} \right)
     \right.
\nonumber \\
     &&~~~~~~~~~~~~   
     \left.
         + ( |\xi_b|^2 + |\xi_s|^2 ) \frac{m_t^2}{m_b^2} 
         \left( 0.112 \frac{F^2_R(x_t)}{F^2(x_t)} 
              + 0.002 \frac{G^2_R(x_t)}{G^2(x_t)} 
              + 0.025 \frac{F_R(x_t) G_R(x_t)}{F(x_t) G(x_t)} \right)
     \right],
\ee
The SM branching ratio is predicted to be
$ {\rm Br}(B \to X_s \gamma) = (3.15 \pm 0.23) \times 10^{-4} $
for $E_\gamma > 1.6$ GeV at next-to-next-to-leading order (NNLO) 
\cite{bsgammaSM}.
The current world average value
of the measured branching ratio is given by
\cite{bsgammaEXP}
\be
{\rm Br}(B \to X_s \gamma) = (3.55 \pm 0.24 ^{+0.09}_{-0.10} \pm 0.03) 
                             \times 10^{-4},
\ee
with the same photon energy cut. 
The allowed parameter sets of $(|\xi_s|, |V_{ts}^{\rm eff}|)$
are depicted in Fig. 1 by green (grey) area
at 95\% C.L..



\section{$\bs$ mixing}

\begin{center}
\begin{figure}[t]
\hbox to\textwidth{\hss\epsfig{file=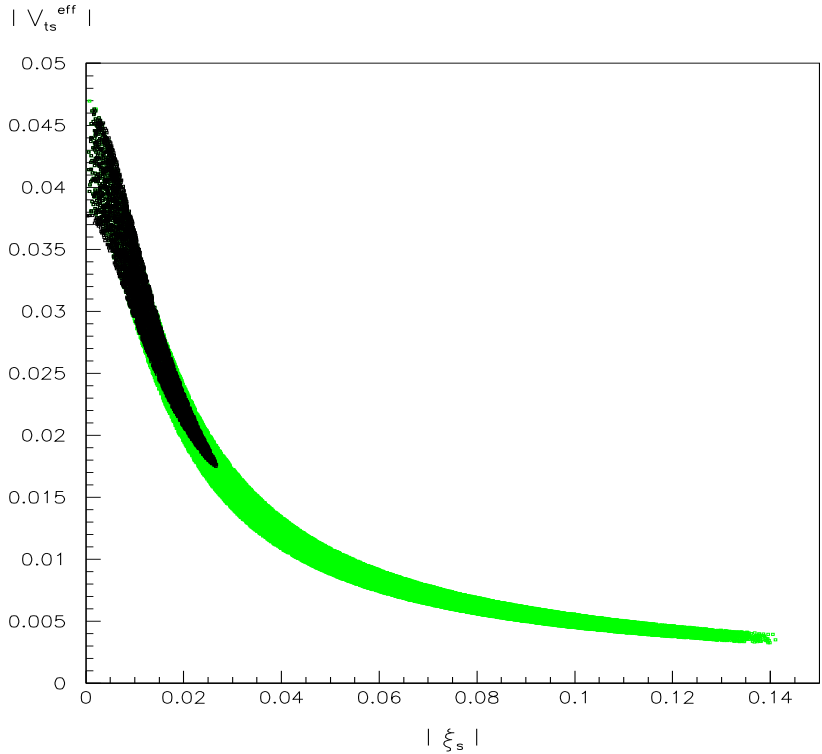,width=13cm,height=11cm}\hss}
 \vskip -1.5cm
\vspace{1cm}
\caption{
Allowed parammeter sets $(|\xi_s|, |V_{ts}^{\rm eff}|)$
constrained by $B \to X_s \gamma$ (green) and by both $\Delta M_s$
and $B \to X_s \gamma$ (black).
}
\end{figure}
\end{center}

A $B_s^0$ meson can oscillate into its antiparticle $\bar{B}_s^0$
via flavour-changing processes of $\bs$ mixing.
The oscillation is represented by the mass difference
between the heavy and light $B_s$ states,
\be 
\Delta M_s \equiv M_H^{B_s} - M_L^{B_s} = 2 | M_{12}^s |,
\ee
where the $\Delta B = 2$ transition amplitudes given by
\be
\langle B_s^0 | {\cal H}_{\rm eff}^{\Delta B = 2} | \bar{B}_s^0 \rangle 
         = M_{12}^s,
\ee
is obtained by the box diagrams
with internal lines of $W$ boson and up-type quarks in the SM.
The new contributions to $\bs$ mixing 
with anomalous top quark couplings given in Eq. (1) 
would be examined with the $\bs$ mixing data.
The $\bs$ mixing is also described by the width difference of the mass
eigenstates
\be
\Delta \Gamma_s \equiv \Gamma_L^s - \Gamma_H^s 
                = 2~ {\rm Re} \frac{\Gamma_{12}^s}{M^s_{12}},
\ee
where the decay widths $\Gamma_L$ and $\Gamma_H$ are corresponding to
the physical eigenstates $B_L$ and $B_H$.
Since the decay matrix elements $\Gamma_{12}^s$ is derived from
the SM decays $b \to c \bar{c} s$ at tree level,
it is hardly affected by the new physics.
We consider the new effects of the anomalous top couplings only in $M_{12}^s$.
Since $\xi_q$ are complex parameters, the new physics effects 
arise in both magnitude and phase of $M_{12}^s$ in general.
In this analysis, 
we just consider the mass difference.
Effects of the phase and CP violation in $M_{12}^s$ 
have been measured \cite{cpexp}, although not very accurately, 
and discussed in several literatures 
\cite{cp}.

Including the odd number of right-handed couplings in the box diagram
does not contribute to the transitin amplitude $M_{12}^s$ 
due to vanishing the loop integral of the odd number of momentum.
Thus the leading contribution of the anomalous right-handed top couplings
to the $\bs$ mixing is quadratic order of $\xi_q$.
Calculating box diagrams including the anomalous couplings,
the transition amplitude is given by
\be
M_{12}^s &=&  \frac{G_F^2 m_W^2}{12 \pi} m_{B_s} 
                 \eta_{B} \hat{B}_{B_s} f_{B_s}^2 S_0(x_t)
     \left( \frac{|{V_{ts}^{\rm eff}}^* V_{tb}^{\rm eff}|}{0.0404} \right)^2
\nonumber \\
           && \times  \left( 1 + \frac{S_3(x_t)}{S_0(x_t)}
                             \left(
 \frac{\xi_s^2}{4} \frac{(\bar{b} P_R s) (\bar{b} P_R s)}
                        {(\bar{b} \gamma^\mu P_L s) (\bar{b} \gamma_\mu P_L s)}
+\frac{\xi_b^\ast \xi_s}{2} \frac{(\bar{b} P_L s) (\bar{b} P_R s)}
                        {(\bar{b} \gamma^\mu P_L s) (\bar{b} \gamma_\mu P_L s)}
+\frac{{\xi_b^\ast}^2}{4} \frac{(\bar{b} P_L s) (\bar{b} P_L s)}
                        {(\bar{b} \gamma^\mu P_L s) (\bar{b} \gamma_\mu P_L s)}
                             \right)
\right),
\ee
where $\eta_B$ is the perturbative QCD correction 
to the $B-\bar{B}$ mixing \cite{QCD}.
The Inami-Lim loop functions are given by
\be
S_0(x) &=& \frac{4x-11 x^2 + x^3}{4 (1-x)^2} - \frac{3 x^3}{2 (1-x)^3} \log x,
\nonumber \\
S_3(x) &=& 4 x^2 \left( \frac{2}{(1-x)^2} + \frac{1+x}{(1-x)^3} \log x \right).
\ee
Using the vacuum insertions, we calculate
\be
\frac{\langle B_s^0 |(\bar{b} P_R s) (\bar{b} P_R s)|\bar{B}_s^0 \rangle}
         {\langle B_s^0 |(\bar{b} \gamma^\mu P_L s) 
                         (\bar{b} \gamma_\mu P_L s)|\bar{B}_s^0 \rangle}
        &=& \frac{5}{8} \left( \frac{m_{B_s}}{m_b+m_s} \right)^2 ,
\nonumber \\
\frac{\langle B_s^0 |(\bar{b} P_L s) (\bar{b} P_R s)|\bar{B}_s^0 \rangle}
         {\langle B_s^0 |(\bar{b} \gamma^\mu P_L s)
                         (\bar{b} \gamma_\mu P_L s)|\bar{B}_s^0 \rangle}
        &=& \frac{3}{4} \left( \frac{1}{6} 
                 - \left( \frac{m_{B_s}}{m_b+m_s} \right)^2 \right) ,
\nonumber \\
\frac{\langle B_s^0 |(\bar{b} P_L s) (\bar{b} P_L s)|\bar{B}_s^0 \rangle}
         {\langle B_s^0 |(\bar{b} \gamma^\mu P_L s) 
                         (\bar{b} \gamma_\mu P_L s)|\bar{B}_s^0 \rangle}
&=&
\frac{\langle B_s^0 |(\bar{b} P_R s) (\bar{b} P_R s)|\bar{B}_s^0 \rangle}
         {\langle B_s^0 |(\bar{b} \gamma^\mu P_L s) 
                         (\bar{b} \gamma_\mu P_L s)|\bar{B}_s^0 \rangle},
\ee
and 
\be
\langle B_s^0 | (\bar{b} \gamma^\mu P_L s) (\bar{b} \gamma_\mu P_L s)
 | \bar{B}_s^0 \rangle = \frac{8}{3} m_{B_s}^2 \hat{B}_{B_s} f_{B_s}^2 ,
\ee
where $\hat{B}_{B_s}$ is the Bag parameter and $f_{B_s}^2$ the decay constant.

We show the allowed parameter sets $(|\xi_s|, V_{ts}^{\rm eff})$ in Fig. 1
by black area at 95\% C.L..
We use the SM prediction $\Delta m_s = 19.3 \pm 6.74$ ps$^{-1}$ 
given in Ref. \cite{nierste}.
The conservative bounds $|\xi_s| < 0.027$ 
and $|V_{ts}^{\rm eff}| > 0.017$ are obtained from this analysis.
The correlated results between observables, 
Br($B \to X_s \gamma$) and $\Delta M_s$
are shown in Fig. 2 with allowed parameters given in Fig. 1 (black area).

\begin{center}
\begin{figure}[t]
\hbox to\textwidth{\hss\epsfig{file=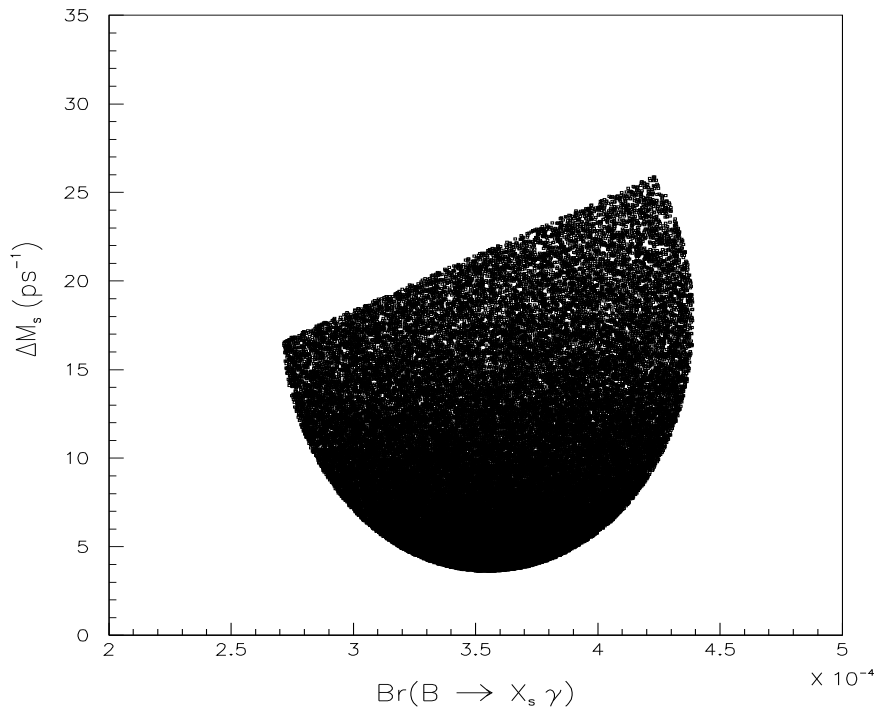,width=13cm,height=11cm}\hss}
 \vskip -1.5cm
\vspace{1cm}
\caption{
Correlation of Br($B \to X_s \gamma$) and $\Delta M_s$
with allowed values of $(|\xi_s|, |V_{ts}^{\rm eff}|)$.
}
\end{figure}
\end{center}

\section{Top quark decays}

The flavour-diagonal $t \to b W$ decay dominates,
${\rm Br}( t \to s W) \approx 1$.
The branching ratio of the CKM-suppressed decays are given by
\be
{\rm Br}( t \to s W) = |V_{ts}^{\rm eff}|^2 (1 + |\xi_s|^2).
\ee
Since there is no enhancement factor involved, 
the branching ratio is insensitive to $\xi_s$
and determined by $V_{ts}^{\rm eff}$.
The predictions of $ {\rm Br}( t \to s W) $ 
is depicted in Fig. 3 with respect to the allowed values of $\xi_s|$.
We find that large deviation of $ {\rm Br}( t \to s W) $ 
from the SM prediction is possible.
The correlation between $\Delta M_s$ and $ {\rm Br}( t \to s W) $ 
are shown in Fig. 4 with allowed parameters given in Fig. 1 (black area).
Both observables of $\Delta M_s$ and $ {\rm Br}( t \to s W) $ 
crucially depend on $V_{ts}^{\rm eff}$
but are insensitive to $\xi_s$.
Since the value of $V_{ts}^{\rm eff}$ will be strongly constrained 
by $ {\rm Br}( t \to s W) $,
the right-handed coupling $\xi_s$ will be also constrained
through $B \to X_s \gamma$ decay
if we measure the branching ratio of $ t \to s W $
at the LHC or the future colliders.

\begin{center}
\begin{figure}[t]
\hbox to\textwidth{\hss\epsfig{file=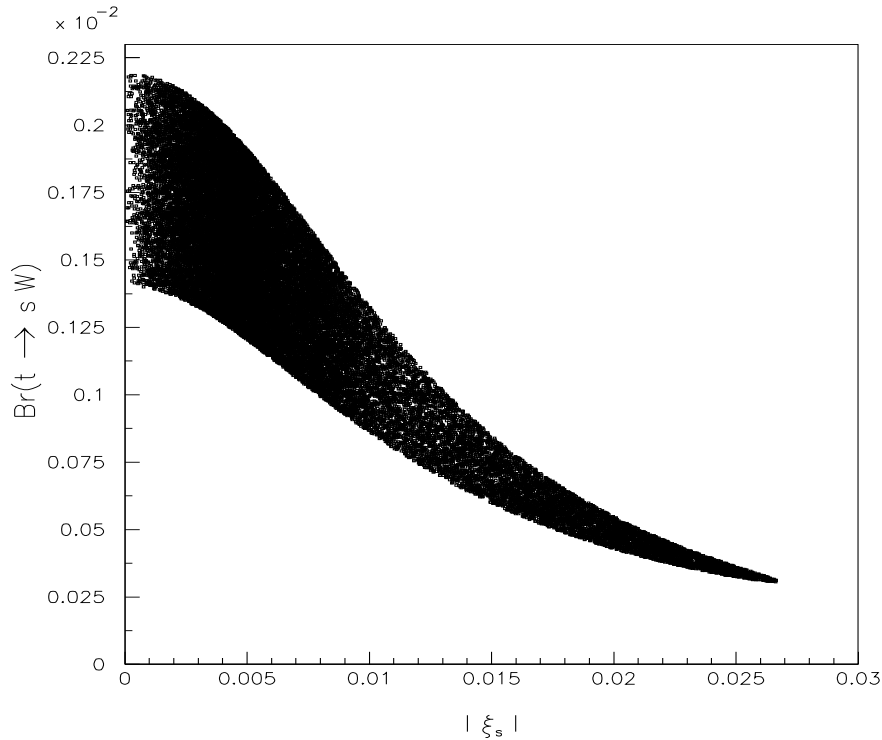,width=13cm,height=11cm}\hss}
 \vskip -1.5cm
\vspace{1cm}
\caption{
Prediction of Br($ t \to s W$) with respect to $|\xi_s|$.
}
\end{figure}
\end{center}

\begin{center}
\begin{figure}[h]
\hbox to\textwidth{\hss\epsfig{file=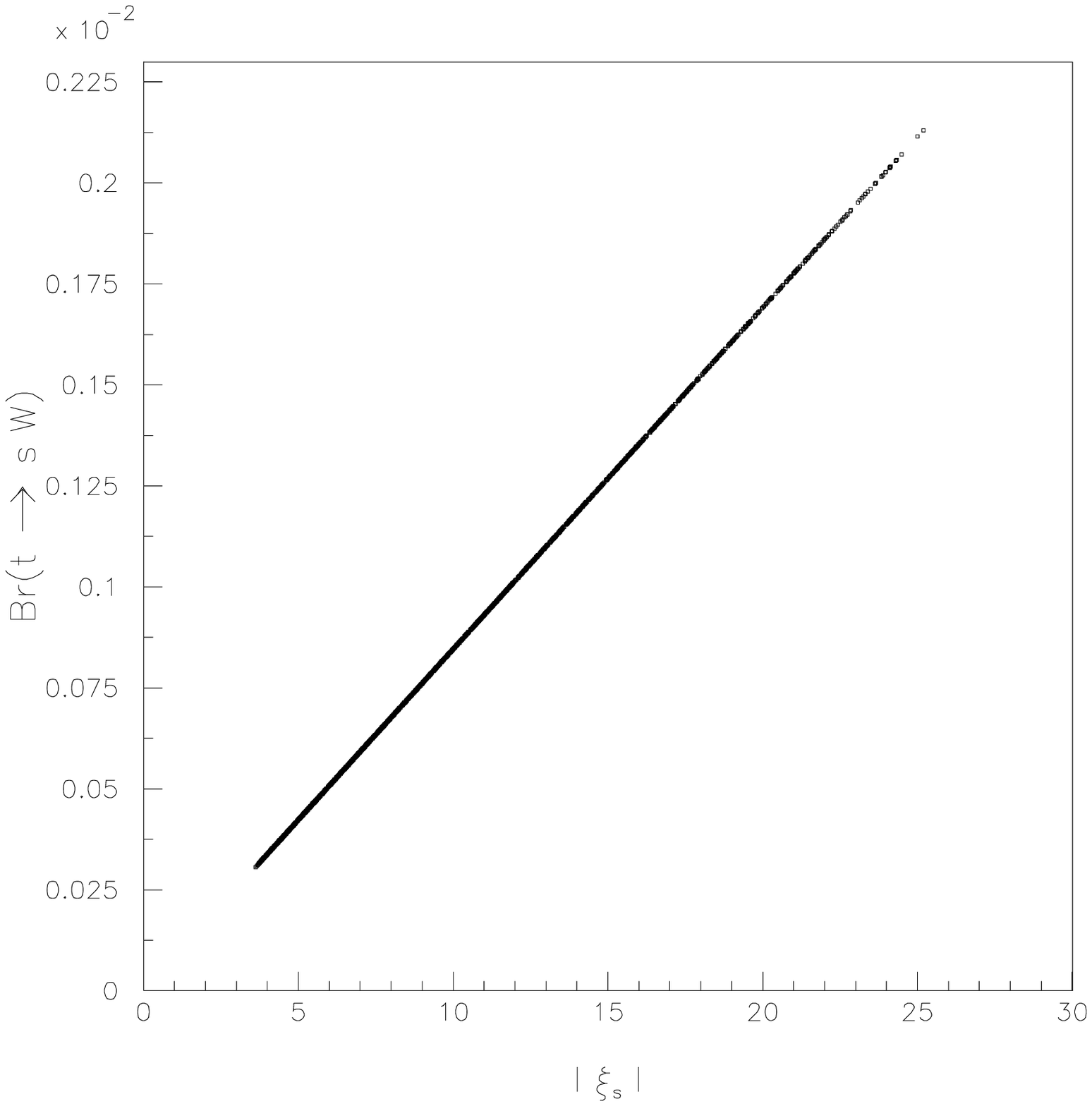,width=13cm,height=11cm}\hss}
 \vskip -1.5cm
\vspace{1cm}
\caption{
Correlation of $\Delta M_s$ and Br($ t \to s W$)
with allowed values of $(|\xi_s|, |V_{ts}^{\rm eff}|)$.
}
\end{figure}
\end{center}

\section{Concluding Remarks}

We consider the anomalous top quark coupling
which are not direct measured yet.
The $\bar{t} s W$ coupling is parametrized
by $V_{ts}^{\rm eff}$ and $\xi_s$.
Combined analysis of $\bs$ mixing and $B \to X_s \gamma$ decay
gives strong constraints on by $V_{ts}^{\rm eff}$ and $\xi_s$.
The prediction of the branching ratio of the top decay
$ {\rm Br}( t \to s W) $ is given 
and it is shown that both of $\Delta M_s$ and $ {\rm Br}( t \to s W) $ 
depend only on $V_{ts}^{\rm eff}$.
In conclusion, we can examine the anomalous $\bar{t} s W$ coupling
through $\bs$ mixing and $B \to X_s \gamma$ decay
and will test it more by the $ t \to s W $ decay
in the future colliders.

\acknowledgments
This work was supported by
the Korea Research Foundation Grant funded by the Korean Government
(MOEHRD, Basic Research Promotion Fund KRF-2007-C00145)
and the BK21 program of Ministry of Education (K.Y.L.).

\def\PRD #1 #2 #3 {Phys. Rev. D {\bf#1},\ #2 (#3)}
\def\PRL #1 #2 #3 {Phys. Rev. Lett. {\bf#1},\ #2 (#3)}
\def\PLB #1 #2 #3 {Phys. Lett. B {\bf#1},\ #2 (#3)}
\def\NPB #1 #2 #3 {Nucl. Phys. {\bf B#1},\ #2 (#3)}
\def\ZPC #1 #2 #3 {Z. Phys. C {\bf#1},\ #2 (#3)}
\def\EPJ #1 #2 #3 {Euro. Phys. J. C {\bf#1},\ #2 (#3)}
\def\JHEP #1 #2 #3 {JHEP {\bf#1},\ #2 (#3)}
\def\IJMP #1 #2 #3 {Int. J. Mod. Phys. A {\bf#1},\ #2 (#3)}
\def\MPL #1 #2 #3 {Mod. Phys. Lett. A {\bf#1},\ #2 (#3)}
\def\PTP #1 #2 #3 {Prog. Theor. Phys. {\bf#1},\ #2 (#3)}
\def\PR #1 #2 #3 {Phys. Rep. {\bf#1},\ #2 (#3)}
\def\RMP #1 #2 #3 {Rev. Mod. Phys. {\bf#1},\ #2 (#3)}
\def\PRold #1 #2 #3 {Phys. Rev. {\bf#1},\ #2 (#3)}
\def\IBID #1 #2 #3 {{\it ibid.} {\bf#1},\ #2 (#3)}


\begin{thebibliography}{99}

\bibitem{top} 
F. Abe {\it et al.}, CDF Collaboration,
\PRL 73 225 1994 ; \PRL 74 2626 1995 ;
\PRL 80 2767 1998 ; \PRL 80 2773 1998 ;
S. Abachi {\it et al.}, D0 Collaboration,
\PRL 74 2632 1995 ; \PRL 79 1197 1997 ; \PRL 79 1203 1997 ;
D. Abbott {\it et al.}, D0 Collaboration,
\PRL 80 2063 1998 ; \PRL 82 271 1999 ;
\PRD 58 052001 1998 ;
\PRD 60 052001 1999 ;
{\it ibid.} 2779 .

\bibitem{lhc} M. Beneke {\it et al.}, hep-ph/0003033.

\bibitem{toplhc} W. Bernreuther, hep-ph/0805.1333.

\bibitem{larios} F. Larios, M.A. Perez, and C.P. Yuan, 
\PLB 457 334 1999 .

\bibitem{leesong} K. Y. Lee and W. Y. Song, \PRD 66 057901 2002 ; 
Nucl. Phys. Proc. Suppl. {\bf 111}, 288 (2002).

\bibitem{lee2} J. P. Lee and K. Y. Lee, \EPJ 29 373 2003 .

\bibitem{lee} K. Y. Lee, \PLB 632 99 2006 . 

\bibitem{jplee} J. P. Lee, \PRD 69 014017 2004 .

\bibitem{boos} E. Boos, A. Pukhov, M. Sachwitz, and H.J. Schreiber, 
\PLB 404 119 1997 ;
E. Boos, M. Dubinin, M. Sachwitz, and H.J. Schreiber,
\EPJ 16 269 2000 .

\bibitem{rindani} S.D. Rindani, Pramana {\bf 54}, 791 (2000).

\bibitem{elhady} A. Abd El-Hady and G. Valencia,
\PLB 414 173 1997 .

\bibitem{yue} C.-X. Yue, G.-R. Lu, and W.-B. Li, 
Chinese Phys. Lett. {\bf 18}, 349 (2001). 

\bibitem{rizzo} T. G. Rizzo, \PRD 58 055009 1998 .

\bibitem{cdf} A. Abulencia {\it et al.}, CDF collaboration, 
\PRL 97 242003 2006 [arXiv: hep-ex/0609040].

\bibitem{d0} D0 collaboration,

\bibitem{buras} A.J. Buras, hep-ph/9806471;
G. Buchalla, A.J. Buras, and M.E. Lautenbacher, 
Rev. Mod. Phys. {\bf 68}, 1125 (1996). 

\bibitem{inami} T. Inami and C.S. Lim, Prog. Theo. Phys. {\bf 65}, 297 (1981).

\bibitem{cho} P. Cho and M. Misiak, \PRD 49 5894 1994 .

\bibitem{kagan} A.L. Kagan and M. Neubert, \EPJ 7 5 1999 .

\bibitem{bsgammaSM} M. Misiak and M. Steinhauser, \NPB 764 62 2007 ;
M. Misiak {\it et al}., \PRL 98 022002 2007 .

\bibitem{bsgammaEXP} Heavy Flavor Averaging Group,
arXiv : hep-ex/0505100,
http://www.slac.stanford.edu/xorg/hfag/.

\bibitem{cpexp} 
V. M. Abazov {\it et al.}, D0 Collaboration, \PRD 76 057101 2007 ;
T. Aaltonen {\it at al.}, CDF Collaboration, arXiv:0712.2397 [hep-ex];
V. M. Abazov {\it et al.}, D0 Collaboration, arXiv:0802.2255 [hep-ex].

\bibitem{cp} 
U. Nierste, \IJMP 22 5986 2008 ;
P. Ball, hep-ph/0703214;
M. Blanke, A. J. Buras, S. Recksiegel and C. Tarantino, 
arXiv:0805.4393[hep-ph];
J. Hisano and Y. Shimizu, arXiv:0805.3327[hep-ph];
P. Ko, Nucl. Phys. Proc. Suppl. {\bf 163}, 185 (2007);
J. K. Parry and H. h. Zhang, arXiv:0710.5443 [hep-ph];
F. J. Botella, G. C. Branco and M. Nebot, arXiv:0805.3995 [hep-ph].

\bibitem{QCD} A. J. Buras, M. Jamin and P. H. Weisz, \NPB 347 491 1990 . 

\bibitem{nierste} A. Lenz and U. Nierste, \JHEP 06 072 2007 .


\end{thebibliography}
\end{document}